\documentstyle[prl,aps,amsmath,epsfig]{revtex}

\begin{document}

\draft 

\title{Simple strong glass forming models: mean--field solution with
activation}

\author{Arnaud Buhot$^{1,2}$\thanks{Email:
buhot@drfmc.ceng.cea.fr.}, Juan P. Garrahan$^1$\thanks{Email:
j.garrahan1@physics.ox.ac.uk.} and David Sherrington$^1$\thanks{Email:
d.sherrington1@physics.ox.ac.uk.}}

\address{$^1$Theoretical Physics, University of Oxford, 1 Keble Road,
Oxford, OX1 3NP, U.K. \\
$^2$CEA Grenoble, DRFMC,
SI3M, 17 rue des Martyrs, 38054 Grenoble, France.}

\date{\today}

\maketitle

\begin{abstract}
We introduce simple models, inspired by previous models for froths and
covalent glasses, with trivial equilibrium properties but dynamical
behaviour characteristic of strong glass forming systems. These models
are also a generalization of backgammon or urn models to a
non--constant number of particles, where entropic barriers are
replaced by energy barriers, allowing for the existence of activated
processes.  We formulate a mean--field version of the models, which
keeps most of the features of the finite dimensional ones, and solve
analytically the out--of--equilibrium dynamics in the low temperature
regime where activation plays an essential role.
\end{abstract}


\section{Introduction}

Glass forming systems, like supercooled liquids, display remarkable
dynamical properties, and are the subject of intense experimental and
theoretical studies (for reviews see \cite{Angell,Ediger,Debe}). Out
of all glass formers, the ones which have the seemingly simpler
macroscopic behaviour are strong liquids, typically covalently bonded
systems like SiO$_2$ and GeO$_2$, in which viscosities and relaxation
times increase exponentially with decreasing temperature, following a
simple Arrhenius law \cite{Angell}. While the increase of timescales
in strong systems is not as dramatic as in fragile ones, it is still
very pronounced, and, moreover, it is not accompanied by the growth of
static lengthscales or similar corresponding structural signatures.

The separation of statics and dynamics is a central feature of
structural glass formers \cite{Krauth}, and is in marked contrast with
other dynamically arrested systems, like spin glasses and other
systems with quenched disorder \cite{Fischer,Young}. This leads
naturally to an approach for modeling these systems based on the idea
that glassiness is not a consequence of disorder or frustration in the
static interactions but of constraints on their dynamics
\cite{Palmer,Fredrickson,constraints,barcelona}. The aim of this paper
is to build on our earlier efforts in the modeling of covalent systems
and froths \cite{Aste,Lexie,Lexie2}, where the dynamical constraints
that lead to glassy slowdown can be related directly to elementary
structural processes, and develop a class of models which retain the
essential features but are simple enough to allow for analytical
solutions, and study in detail the connection with underlying
effective diffusion--annihilation processes \cite{Cardy,Hinrichsen}
which we believe are an essential component of strong glass formers in
general.
 
The models we study in this work are described explicitly and
concisely in section \ref{model}. Here we outline briefly the
background considerations which were their conceptual basis. They are
a distillation of a series of minimalist models exhibiting strong
glass--like dynamical behaviour through the operation of kinetic
constraints despite non--interacting Hamiltonians and trivial
thermodynamics (with no equilibrium phase transitions) and no imposed
quenched disorder.

\begin{center}
\begin{figure}[bt]
\epsfig{figure=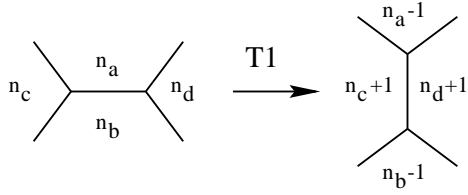,height=6.5cm,angle=-90}
\caption{\label{T1} A $T1$ move. The initially adjacent cells 
decrease their number of neighbours by $1$ whereas the cells 
which become adjacent increase their number of neighbours by $1$.} 
\end{figure}
\end{center}

The first model\cite{Aste} was a 2--dimensional topological `foam'
consisting of a fully pairwise connected network of three--armed
vertices with energy function $E = \sum_i (n_i - 6)^2$ where $n_i$ is
the number of edges (sides) of cell $i$ ; thus it emulates desire for
local hexagonal structure without the complication of interaction.
The dynamics is $T1$ (see Fig.~\ref{T1}) performed stochastically with
acceptance probabilities determined by the Boltzmann ratio
$\exp(-\Delta E/T)$ where $\Delta E$ is the resultant energy change
and $T$ is the temperature. This ensures eventual equilibration at any
finite temperature but also provides a kinetic constraint since each
$T1$ move involves four cells, decreases $n$ by $1$ for each of the
two initially adjacent cells and increases $n$ of the other two by the
same amount.

This model can be considered variously as an idealization of a glass
of covalent $sp^2$ hybrid bonds which change connections, or of the
Voronoi cells surrounding a fluid of atoms, or of a foam ; in the
first case effectively allowing for harmonic changes of vertex angle
energy but ignoring cell--cell correlation energies, in the second and
third replacing continuous energies associated with interatomic forces
or surface tension by the simple topological energies. At low
temperatures the model exhibits two--timescale macro--behaviour, for
example in the decay of the energy from a random start or in the
autocorrelation function even in equilibrium. The fast timescale is
temperature independent and the slow timescale is Arrhenius--like:
$\tau \sim \exp (a/T)$, $a\simeq 2$\cite{Aste,Lexie}. This behaviour
can be understood by considering $n_i = 6$ as `perfect' and $n_i \neq
6$ as `defect' cells, and the cooperative behaviour in terms of the
diffusion and annihilation of defects. The fast processes correspond
to ones which involve either an energy decrease or no energy change,
while the slow processes involve energetic defect creation. Defects
with $|n_i - 6| > 1$ disappear rapidly at low temperature so we
concentrate on $|n_i-6|\leq 1$. Let us denote $n = 5 (7)$ cells by
$A(\bar{A})$ and $n=6$ by $\emptyset$. Then energy reduction is due to
the annihilation of an $A\bar{A}$ neighbouring pair (or dimer). The
$T1$ process however involves four cells and only allows $A\bar{A}$
dimer annihilation if at most only one of the other cells involved is
a $\emptyset$; thus the annihilation of two dimers\cite{footnote1}
\begin{equation}
\label{2anni}
2 A + 2 \bar{A} \to 4 \emptyset
\end{equation}
exists as well as the single dimer annihilation\cite{footnote2}
\begin{equation}
\label{anni}
2 A + \bar{A} + \emptyset \to 3 \emptyset + A, \;\;\;
2 \bar{A} + A + \emptyset \to 3 \emptyset + \bar{A} .
\end{equation}
$A\bar{A}$ dimers can diffuse on the fast (microscopic attempt)
timescale if the other pair of cells in the $T1$ process are both
$\emptyset$:
\begin{equation}
\label{diffusion}
A + \bar{A} + 2 \emptyset \to 2 \emptyset + A + \bar{A}.
\end{equation}
Defects are only annihilated in dimer pairs ($A\bar{A}$) and isolated
defects must be paired to annihilate, hence they must be moved
together. However they can only move via processes which involve the
creation of an $A\bar{A}$ dimer and a consequent energetic penalty;
i.e. by the inverses of processes (\ref{anni})
\begin{equation}
\label{creation}
A + 3 \emptyset \to \emptyset + 2 A + \bar{A} , \;\;\;
\bar{A} + 3 \emptyset \to \emptyset + 2 \bar{A} + A ,
\end{equation}
which are slower by the Boltzmann factor $\exp (-2/T)$. Dimer pairs,
of either of the two possible $A\bar{A}$ combinations on the right
hand sides of processes (\ref{creation}) can then diffuse away freely
by process (\ref{diffusion}). Processes (\ref{anni}) themselves also
allow slow movement of $A$ (or $\bar{A}$) via collision with a roaming
$A\bar{A}$ dimer, leading to annihilation of all three $A, \bar{A}$
but creation of $A$ (or $\bar{A}$) on the cell which was originally
$\emptyset$. This also involves a slowing factor $\exp(-2/T)$ since
the equilibrium density of $A\bar{A}$ dimers involves such a
factor. As a consequence, at low temperatures, the energy after a
quench from a random/high temperature start quickly decays to a
plateau through elimination of $A\bar{A}$ dimers, leaving mainly (and
at $T= 0$ only) isolated defects and clusters of only like
defects. After this plateau, the energy further decays to equilibrium
with the Arrhenius--like timescale $\tau \sim \exp(a/T)$ due to the
effective dimer--mediated diffusion of isolated defects and eventual
$A\bar{A}$ pairing and annihilation.

The second generation of models is built on the above picture,
replacing the foam by lattice--based models with `spins' associated
with the dual plaquettes\cite{footnote3}
\begin{equation}
S_i = 1,0,-1,
\end{equation}
the energy by
\begin{equation}
E = J \sum_{i=1}^N S_i^2,
\end{equation}
and the $T1$ move by one involving four spins around a corresponding
`Feyman diagram' edge in a hexagonal lattice, a square vertex in a
square lattice, etc.\cite{Lexie2}. For $J > 0$ this behaves
essentially as for the foam model\cite{footnote4}, but allows much
larger simulations. The resultant better--accuracy data permits
comparison with the asymptotic predictions of the field theory of
simple diffusion--annihilation processes\cite{Cardy,Hinrichsen} of
type $A + B \to \emptyset$ (in the usual notation of that subject);
the agreement is quite good for both decays which can be viewed as
diffusion--annihilation processes with different
timescales\cite{Lexie2}.
 
The second class of models permits a further modification, the taking
of $J<0$. In this case the ground state is highly degenerate ($S_i=\pm
1$ independently on each plaquette), as compared with the unique
ground state for $J>0$ ($S_i=0$; all $i$), but the defects are of only
one type ($S_i=0$). The two timescale behaviour of the energy decay to
equilibrium and of the equilibrium autocorrelation functions persists.
Ignoring the complications of the ground state degeneracy and further
constraints imposed by the $T1$ process ($|S_i|$ cannot be increased
above $1$) the behaviour is now describable in terms of only $A$,
$\emptyset$ dynamics (the $\emptyset$ now referring to $S=\pm 1$ and
$A$ to $S=0$). The qualitative discussion earlier still applies with
$\bar{A}$ (or $B$) replaced by $A$. Again the comparison with the
predictions of simple field theory of diffusion--annihilation is fair,
but less precise, presumably in part due to the more complicated
ground state degeneracy and constraints.

The models studied in this paper are a further simplification which
maintain the features of (i) fast annihilation of dimers, (ii) fast
diffusion of dimers, (iii) motion of isolated defects only by slow
creation of dimers. It allows for either a single defect type ($A$)
and identical--defect dimers ($AA$) or two defect types ($A$ and $B$)
and mixed dimers ($AB$), but in both cases with a non--degenerate
($T=0$ absorbing) ground state ($\emptyset$). They also allow
separation of the timescales for processes (i) and (ii) above, as well
as their separation from (iii). The explicit formulation is given in
the next section, but the above should clarify why they are
appropriate.

The rest of the paper is organized as follows. In section \ref{model}
we give a full description of the models studied in this work.  Their
general features are discussed and compared with those of previous
models is section \ref{features}. In section \ref{solution} we solve
analytically the mean--field version of the models. Our conclusions
are given in section \ref{conclusions}. Details of the analytical
calculations are provided in the appendices.

\section{Description of the models}
\label{model}

The minimalist models we study here are based on a coarse--grained
simplification of the ideas described above. They correspond to
defects (or particles), which can be either of a single kind $A$, or
different kinds, $A$ and $B$, and live in a $d$ dimensional lattice.
They can also be considered as a generalization of backgammon
\cite{Ritort} or urn models \cite{Luck} to a non--constant number of
particles, with energetic barriers rather than entropic ones, thus
allowing for the existence of activated processes. The analogy with
previous models \cite{Aste,Lexie,Lexie2} is obtained through the
dynamical rules which mimic the different processes (\ref{anni}),
(\ref{diffusion}) and (\ref{creation}). The two models under
consideration are the following.

\subsection{Single type of particles}

As for previous models, with each defect (particle in our case) we
associate a unit energy ($J = 1$) leading to the Hamiltonian
\begin{equation}
H = \sum_{i=1}^{N} n_i ,
\label{hache}
\end{equation} 
with $n_i$ the occupation number on site $i$. Due to the
non-interacting Hamiltonian, the equilibrium properties are
trivial. The probabilities $p^{eq}_n$ to have $n$ particles on a given
site at temperature $T = 1/\beta$ are:
\begin{equation}
\label{proba}
p_n^{eq}(\beta) = e^{-\beta n} \left( \sum_{n=0}^{n_{\rm max}} 
e^{-\beta n} \right)^{-1}  
\end{equation} 
and lead to an equilibrium energy or concentration of particles:
\begin{equation}
\label{concentration}
c_{eq}(\beta) = \sum_{n=0}^{n_{\rm max}} n p_n^{eq} (\beta) 
\end{equation}
with $n_{\rm max}$ the maximum number of particles per site.  In the
following we use the smallest number ($n_{max} = 3$) compatible with
the dynamical rules. The infinite temperature concentration
$c_{eq}(\beta = 0) = 3/2$ corresponds to equal probabilities $p_n^{eq}
= 1/4$ whereas the low temperature concentration vanishes as
$c_{eq}(\beta) \sim e^{-\beta}$.

The dynamical rules are inspired from the $T1$  
moves\cite{Aste,Lexie,Lexie2}. 
Three different kinds of moves are considered\cite{footnote_2dimer}:\\ 
(i) The annihilation of two particles analogous to processes 
(\ref{anni}): three particles disappear from site $i$ and only 
one appears on a neighbouring site $j$ with a rate $1$ 
\begin{equation}
(n_i,n_j) \to (n_i-3,n_j+1)
\end{equation}
(ii) The dimer diffusion analogous to process (\ref{diffusion}): 
two particles move from site $i$ to a neighbouring site $j$ with 
a diffusive rate $D$ 
\begin{equation}
(n_i,n_j) \to (n_i-2,n_j+2)
\end{equation}
(iii) The creation of two particles analogous to processes 
(\ref{creation}): a particle disappears from site $i$ to create 
three particles on a neighboring site $j$ with a rate $e^{-2 \beta}$  
\begin{equation}
(n_i,n_j) \to (n_i-1,n_j+3)
\end{equation}
Those moves have also to respect the maximal number of particles per
site $n_{max} = 3$. The rates considered satisfy detailed balance with
respect to Hamiltonian (\ref{hache}), ensuring that the equilibrium
properties [equations (\ref{proba}) and (\ref{concentration})] will be
reached asymptotically by the dynamics. All these processes are of the
form $(n_i,n_j) \to (n_i-x,n_j+y)$ with $x+y=4$, which reflects the
four cell character of the original model transitions.

The introduction of a diffusive constant $D$ allows to separate
explicitly the timescales for diffusion of dimers from that of the
annihilation process. It is also interesting to notice that those
rules are defined for any dimension and any lattice, even for a fully
connected network, leading the possibility to study the mean--field
version of the model.

\subsection{Two different types of particles}

A possible modification of the model consists in considering two
different types of particles ($A$ and $B$) to keep the analogy with
the possible sign of the topological charges $q_i = 6 - n_i$ of the
cells in the foam model. As a consequence, the dynamical rules are
closer to the $T1$ move.  This modification allows to stress the
relation between the last decay of the energy or concentration of
particles and diffusion--annihilation
processes\cite{Cardy,Hinrichsen}, since the critical dimension and
exponents for the $A + B \to \emptyset$ process is different from
those of the $A + A \to \emptyset$ one.

The presence of two types of particles introduce modifications to the
equilibrium properties. In the following we still consider a
restriction on the number of particles per site ($n_{\rm max} = 3$)
irrespective of their nature, but, in addition, the difference between
the numbers of $A$ and $B$ particles on a site is limited to $-1, 0$
or $1$ (this last restriction absent within the previous model avoids
configurations with all $A$'s or all $B$'s on a given site).  As a
consequence the equilibrium probabilities to have $n_A$ particles $A$
and $n_B$ particles $B$ on a given site are (with $n = n_A + n_B$):
\begin{equation}
\label{probaAB}
p_{n_A,n_B}^{eq} (\beta) = \frac{1}{Z}e^{-\beta n} \Theta
(n_{\rm max} - n) \Theta (1 - |n_A-n_B|)
\end{equation}
leading to the equilibrium concentration of particles:
\begin{equation}
\label{concentrationAB}
c_{eq} (\beta) = \sum_{n_A,n_B} n p_{n_A,n_B}^{eq} (\beta)
\end{equation}
with:
\begin{equation}
Z = \sum_{n_A,n_B} e^{-\beta n} \Theta(n_{\rm max}-n)
\Theta (1 - |n_A-n_B|)
\end{equation}
and $\Theta(x)$ the Heaviside function ($\Theta(x) = 1$ if $x \geq 0$ 
and $\Theta(x) = 0$ otherwise).

The infinite temperature concentration of particles is $c_{eq} (\beta
= 0) = 5/3$ [six possible configurations with equal probabilities:
$(n_A,n_B) = (0,0), (1,0), (0,1), (1,1), (2,1)$ and $(1,2)$] whereas
for low temperatures the concentration vanishes as $c_{eq}(\beta) \sim
2 e^{-\beta}$ (the coefficient $2$ is coming from an entropic effect
due to the two types of particles).

The dynamical rules are a straightforward generalization of the $T1$
moves\cite{Aste,Lexie,Lexie2}. Three different kinds of moves
are considered:
 
\noindent
(i) annihilation of an $AB$ dimer analogous to processes (\ref{anni}),
an $AB$ dimer and another particle disappear from site $i$ but only
the single particle appears on a neighbouring site $j$ with a rate $1$
\begin{mathletters}
\begin{equation}
[(AAB)_i,(X)_j] \to [(\emptyset)_i,(AX)_j] , \;\;\;
[(ABB)_i,(X)_j] \to [(\emptyset)_i,(BX)_j]
\end{equation}
\end{mathletters}

\noindent
(ii) $AB$ dimer diffusion analogous to process (\ref{diffusion}), a
dimer moves from site $i$ to a neighbouring site $j$ with a diffusive
rate $D$
\begin{equation}
[(ABX)_i,(Y)_j] \to [(X)_i,(ABY)_j]
\end{equation}

\noindent
(iii) creation of an $AB$ dimer analogous to processes
(\ref{creation}), a single particle from site $i$ move to a
neighbouring site $j$ creating a dimer on this site with a rate $e^{-2
\beta}$
\begin{mathletters}
\begin{equation}
[(AX)_i,(\emptyset)_j] \to [(X)_i,(AAB)_j], \;\;\;
[(BX)_i,(\emptyset)_j] \to [(X)_i,(ABB)_j]
\end{equation}
\end{mathletters}
In all of these processes, symbols $X$ and $Y$ stand for possible $A$,
$B$ or $\emptyset$ particles respecting the restrictions in the number
of particles on each site. The rates again satisfy detailed balance
conditions, ensuring equilibration.

\section{General features}
\label{features}

We now show that the two models introduced in the last section share
common behaviour for all dimensions and with the models considered
previously\cite{Aste,Lexie,Lexie2}. In what follows we discuss
equilibrium dynamical properties, in particular the existence of two
different timescales, as well as out--of--equilibrium features like
the multi--stage decay of the energy density after a quench.

\subsection{Dynamics in equilibrium}  

\begin{center}
\begin{figure}[bt]
\epsfig{figure=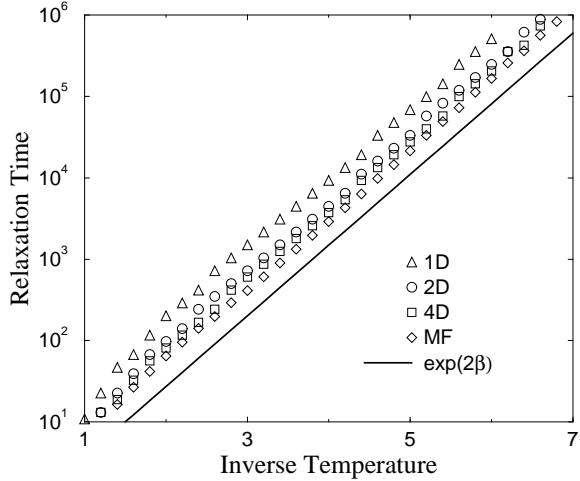,width=6.5cm,angle=-90}
\caption{\label{timel} Relaxation time as function of the inverse
temperature for different dimensions ($d = 1,2,4$ and $\infty$) and a
diffusive constant $D = 10^{-4}$. The line corresponds to the expected
$\tau \propto e^{2\beta}$ behaviour. }
\end{figure}
\end{center}

Let us first consider the equilibrium dynamical properties of these
systems. As we already mentioned, the dynamical behaviour is
characteristic of a strong glass\cite{Angell}. The relaxation time
increases exponentially with the inverse temperature, corresponding to
an Arrhenius law. A simple way to determine this relaxation time is
from the auto--correlation function:
\begin{equation}
C(t,t') = \langle n_i(t) n_i(t') \rangle
\end{equation}
with the brackets denoting ensemble average. In equilibrium this
two--time function reduces to a single time equilibrium correlation
$C_{eq}(t-t')$ due to the time translational invariance. We can define
the relaxation time $\tau$ from $C_{eq}^c(\tau) = C_{eq}^c(0)/e$,
where the connected correlation $C_{eq}^c(t) = C_{eq}(t) -
c_{eq}^2$. At low temperatures and for all diffusive constants, the
temperature dependent and slower process is the creation of particles
which has energy barrier $\Delta E = 2$.  As a consequence we expect
the following Arrhenius law for the relaxation time:
\begin{equation}
\label{arrhenius}
\tau(\beta) \propto e^{2 \beta}.
\end{equation} 
This behaviour is confirmed by numerical simulations.  The expected
Arrhenius law (\ref{arrhenius}) is recovered for all dimensions and
all diffusive constants (see Fig.~\ref{timel}). Similar results are
found for the model with two types of particles. (Simulations have
been performed using continuous time Monte Carlo \cite{Mark} for
systems of $N=10^6$ sites.)

\begin{center}
\begin{figure}[bt]
\epsfig{figure=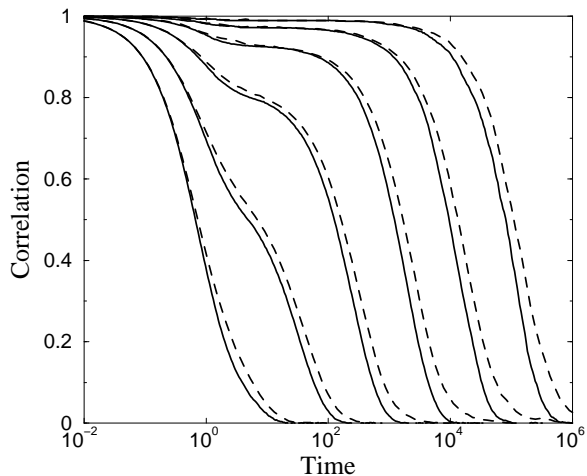,width=6.5cm,angle=-90}
\caption{\label{correleq} Normalized equilibrium autocorrelation
$C^c_{eq}(t)/C^c_{eq}(0)$ for the model with a single type of
particles and a diffusive constant $D = 1$. Different temperatures are
considered (from left to right $\beta = 1,2,3,4,5$ and $6$) as well as
dimensions: $d = 2$ (dashed curves) and mean--field $d = \infty$ (full
curves).}
\end{figure}
\end{center}

The equilibrium correlation explicitly shows the two timescale
behaviour (see Fig.~\ref{correleq}): (i) dimer diffusion or
annihilation on a short timescale independent of the temperature and
(ii) isolated particle motion through the creation of a dimer on the
relaxation timescale.  At low temperatures when the two timescales are
well separated the correlation presents a plateau separating the two
relaxing decays. The relative position of this plateau is increasing
with decreasing the temperature due to the fast disappearance of
dimers.  The decay to zero of the connected correlation is also
dimension dependent.  A change of behaviour occurs between the
mean-field case ($d=\infty$), where it is exponential, and finite
dimensions ($d=2$ in Fig.~\ref{correleq}), where it only vanishes
algebraically. The probability for a particle to come back to the same
place (which depends on the dimension of the system) explains this
difference.

\subsection{Out--of--equilibrium dynamics}

We now consider the out--of--equilibrium behaviour of the models, in
particular the decay of the concentration of particles (energy
density) $c(t) \equiv N^{-1} \langle H(t) \rangle$, after a quench
from an infinite temperature to a low temperature $T$ at time
$t=0$. This decay shows an interesting structure with two intermediate
plateaux when the diffusive constant $D$ and the final temperature $T$
are such that $e^{-2\beta} \ll D \ll 1$.  The first regime is
dominated by the annihilation process and leads to a configuration
with less than three particles on the same site. This first regime
occurs on a timescale of order $1$.  Then, the diffusion process comes
into play on a timescale of order $D^{-1}$ and the dimers diffuse
until they reach a single particle and annihilate. Then the system
reaches a configuration with mainly isolated particles.  Finally, in
order to reach the equilibrium concentration of particles, the
activated regime involving the effective motion of isolated particles
through the creation or annihilation of dimers is necessary and occurs
on a timescale of order $e^{2\beta}$.

The last regime in the concentration decay (before the equilibrium 
concentration is reached) may also be seen as either $A + A \to 
\emptyset$ or $A + B \to \emptyset$ reaction--diffusion processes,
depending on the models, since the particles have to pair themselves 
in order to disappear. Those two processes have different critical 
behaviours and critical dimensions: $d_c = 2$ for the former and 
$d_c = 4$ for the later. As a consequence, we expect a power law 
decay: 
\begin{equation}
c(t) \sim \left(e^{2\beta}/t \right)^{\alpha}
\end{equation}
with $\alpha = 1$ above the critical dimension $d_c$, while below the
critical dimension $d < d_c$, $\alpha = d/2$ for the $A+A \to
\emptyset$ case, and $d/4$ for the $A+B \to \emptyset$ case
\cite{Cardy,Hinrichsen} (at the critical dimension there may be
logarithmic corrections).

\begin{figure}[bt]
\begin{center}
\epsfig{figure=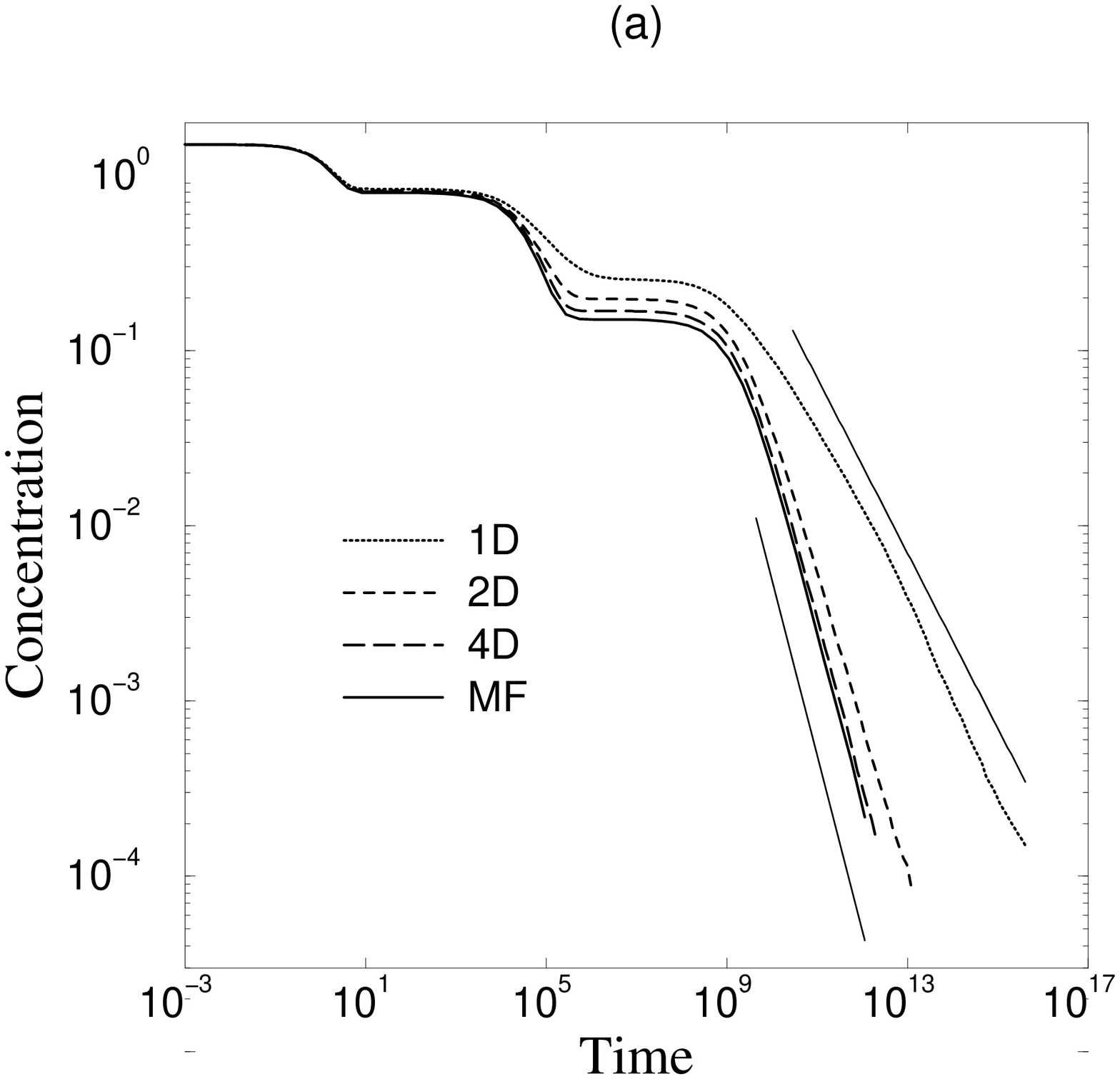,width=6.5cm,angle=-90}
\epsfig{figure=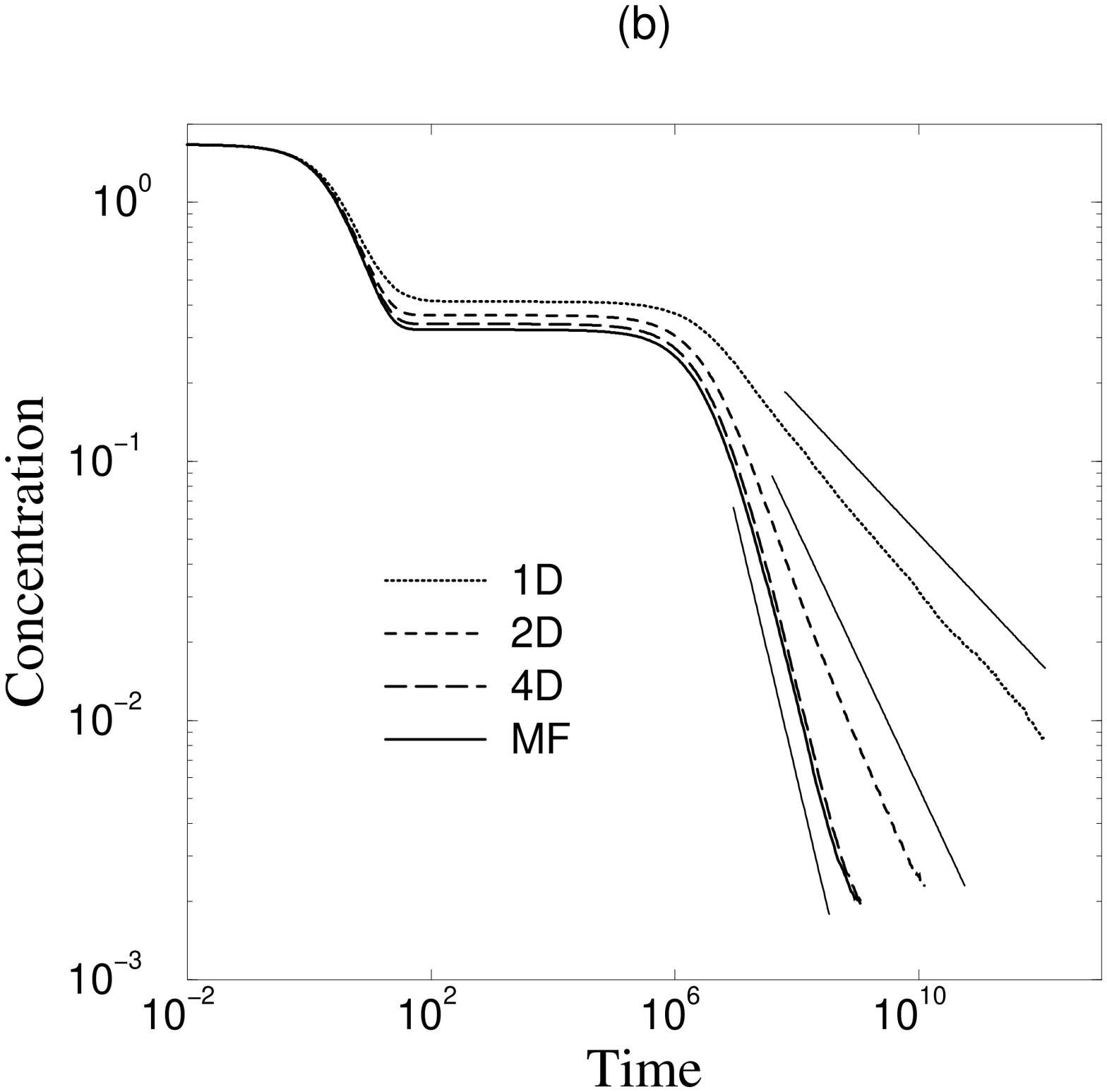,width=6.5cm,angle=-90}
\caption{\label{decay} (Left) Concentration of particles for the model
with a single type of particles after a quench from $T=\infty$ to
$T=1/10$ with a diffusive constant $D = 10^{-4}$ and for different
dimensions ($d = 1,2$ and $4$ and mean--field).  The $A + A \to
\emptyset$ reaction--diffusion process during the second decay is
illustrated by the change in power law for $d < d_c = 2$. The two
straight lines correspond to the expected exponents $\alpha = 1$ and
$\alpha = 1/2$. (Right) Model with two different types of particles,
and diffusive constant $D = 1$.  The straight lines correspond to the
expected power law decays corresponding to $A + B \to \emptyset$
reaction--diffusion processes in different dimensions: for $d < d_c =
4$, $\alpha = d/4$, while $\alpha = 1$ for $d \geq d_c$.}
\end{center}
\end{figure}

In Fig.~\ref{decay} (left) we present numerical simulations for
different dimensions and a diffusive constant $D = 10^{-4}$ of the
concentration decay for the model with a single type of particles.
The temperature after the quench is $T = 1/10$, so the different
timescales are well separated ($1 \ll D^{-1} \ll e^{2\beta}$), and the
decay presents a two plateau structure. The first plateau is roughly
independent of the dimension whereas the second plateau decreases with
increasing dimensions to reach the mean--field value. Notice that the
qualitative behaviour is maintained even in the mean--field limit.  The
dynamics during the last stage of the decay corresponds to a power law
decay with the expected critical exponents $\alpha$.  Fig.~\ref{decay}
(right) shows similar results for the model with two types of
particles, but with a diffusive constant $D = 1$.  The diffusive
timescale is now equivalent to the annihilation one, and the decay
presents a single plateau structure.  Again, we see the critical
behaviour during the last stage of the decay, with the critical
exponents $\alpha$ expected from the theory.

In what follows we will concentrate mainly on the model with one 
kind of particle.

\section{Mean--field solution}
\label{solution}

The mean--field version of the model where all sites are neighbors
allows to write exact evolution equations for the probability
$p^i_n(t)$ of a site $i$ to be occupied by $n$ particles at time $t$.
The dynamical equations for $p^i_n(t)$ read:
\begin{mathletters}
\label{dynamic}
\begin{eqnarray}
\label{dyn_a}
\frac{dp^i_0}{dt} & = & - p^i_0 p_3 + p^i_3 (1-p_3) - D p^i_0
(p_2+p_3)+ D p^i_2 (p_0+p_1) - e^{-2\beta} p^i_0 (1-p_0) + e^{-2\beta}
p^i_1 p_0,\\
\label{dyn_b}
\frac{dp^i_1}{dt} & = & - p^i_1 p_3 + p^i_0 p_3 - D p^i_1 (p_2+p_3) +
D p^i_3 (p_0+p_1) - e^{-2\beta} p^i_1 p_0 + e^{-2\beta} p^i_2 p_0,\\
\label{dyn_c}
\frac{dp^i_2}{dt} & = & - p^i_2 p_3 + p^i_1 p_3 - D p^i_2 (p_0+p_1) +
D p^i_0 (p_2+p_3) - e^{-2\beta} p^i_2 p_0 + e^{-2\beta} p^i_3 p_0,\\
\label{dyn_d}
\frac{dp^i_3}{dt} & = & - p^i_3 (1-p_3) + p^i_2 p_3 - D p^i_3
(p_0+p_1)+ D p^i_1 (p_2+p_3) - e^{-2\beta} p^i_3 p_0 + e^{-2\beta}
p^i_0 (1-p_0).
\end{eqnarray}
\end{mathletters}
The time dependence has been omitted for conciseness and the sum over
the neighbors has been performed using the fact that the probabilities
$p^j_n$ are indeed independent of the site $j$:
\begin{equation}
\frac{1}{N-1} \sum_{j\neq i} p^j_n = p_n = p^i_n.
\end{equation}  
Notice that the right hand size of each of the equations
(\ref{dynamic}) comprises three pairs of terms, each pair
corresponding to a particular process, annihilation, diffusion and
creation, with their respective rates.

The dynamical equations satisfy the conservation of probability,
$\sum_n p^i_n = 1$, which reduces the number of independent variables
to only three, for example $p_0, p_1$ and $p_2$, in the mean--field
model. The trivial equilibrium probabilities:
\begin{equation}
p_n^{eq} = e^{-\beta n} \left(\sum_{k=0}^{3} e^{-\beta k} \right)^{-1} 
\end{equation}
are a stationary solution of Eqs. (\ref{dynamic}). Equations
(\ref{dynamic}) can be solved numerically to arbitrary
accuracy. However, it is also helpful to consider their approximate
solution, regime by regime, to better illustrate the underlying
physics.

\subsection{Concentration decay after a quench}

After a quench from infinite temperature 
to a temperature $T = 1/\beta$, the energy per site or concentration 
of particles $c(t)$ decays from $c(0) = 3/2$ [$p_n(0) = 1/4$ 
for all $n\leq 3$] to its equilibrium value
\begin{equation}
c(\infty) = c_{eq}(\beta) = \left(\sum_{k=0}^{3} k e^{-\beta k}
\right) \left( \sum_{k=0}^{3} e^{-\beta k} \right)^{-1}.
\end{equation}
If $T$ is low and the diffusive constant small, such that $e^{-2
\beta} \ll D \ll 1$, the energy decay may be decomposed into three
different regimes.  The first regime corresponds to the disappearance
of sites with $3$ particles due to the annihilation process and leads
to a first plateau in the decay. The diffusion comes into play in the
second regime and leads to a second plateau.  On this plateau the
particles are essentially isolated.  The last regime is an activated
one where the creation process is necessary to reach the equilibrium
concentration.  In previous models \cite{Aste,Lexie,Lexie2},
where the diffusive timescale of dimers was equivalent to their
annihilation timescale, the first plateau discussed here was absent.

\subsubsection{First regime: zero temperature and no diffusion}

In the first regime, we assume that only the annihilation process
can occur. The equations (\ref{dynamic}) simplify, and only the two 
first terms on the right hand side are relevant. 
The probability $p_3$ decays to $p_3 = 0$, and, assuming that 
$p_2$ stays constant, we get a good approximation for the
different probabilities $p_n(t)$ (See appendix \ref{app_decay} and 
Fig.~\ref{decay1}). The probabilities $p_n$ corresponding to the first 
plateau of the concentration decay are given by:
\begin{equation}
\label{proba1}
\tilde{p}_0 \simeq 0.458, \ \  \tilde{p}_1 \simeq 0.287, 
\ \  \tilde{p}_2 \simeq 1/4, \ \ \tilde{p}_3 \simeq 0.
\end{equation}  

\begin{center}
\begin{figure}[bt]
\centerline{\epsfig{figure=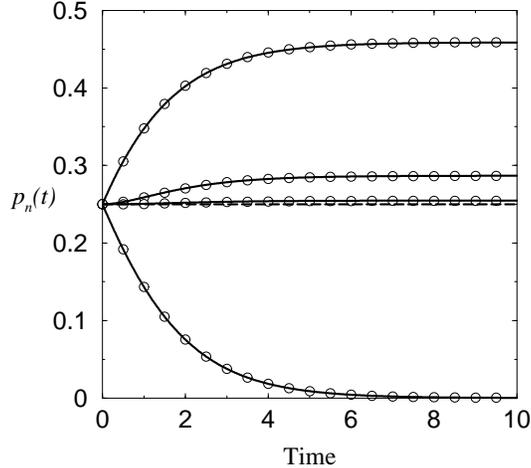,width=6.5cm,angle=-90}}
\caption{\label{decay1} Probabilities $p_n(t)$ during the first regime
of the energy decay. From top to bottom: $p_0, p_1, p_2$ and
$p_3$. Symbols correspond to simulations and lines to the numerical
solution of the exact equations (full lines) and the analytical
approximation (dashed lines). The difference between the two sets of
lines is only visible for $p_2$.}
\end{figure}
\end{center}

\subsubsection{Second regime: zero temperature and slow diffusion}

For slow diffusion, $D \ll 1$, the first regime occurs on a timescale
$t \sim O(1)$. For times $t \sim O(D^{-1})$ diffusion comes into play.
The probability $p_3$ for a site to be occupied by $3$ particles
saturates to a quantity of order $D$. From (\ref{dyn_d}), noticing
that $dp_3/dt \sim O(D^2)$ and thus negligible,
\begin{equation}
\label{p3D}
p_3 \simeq D \frac{p_1 p_2}{p_0+p_1}.
\end{equation} 
From the final probabilities $\tilde{p}_n$ from the first regime,
given by Eq.(\ref{proba1}), it is possible to determine the value of
the second plateau in the concentration decay.  Replacing the
expression (\ref{p3D}) for $p_3$ in Eqs. (\ref{dyn_b}) and
(\ref{dyn_c}), we obtain:
\begin{equation}
\frac{dp_1}{dp_2} = \frac{2 p_1}{1-p_1}
\end{equation}
which can be integrated from $\tilde{p}_2 = 1/4$ to $p_2 (t)$ 
[with $t \sim O(D^{-1}$)]. This leads to  
\begin{equation}
\ln p_1(t) - p_1(t) - \ln \tilde{p}_1 + \tilde{p}_1 = 2 p_2(t) - 
2 \tilde{p}_2.
\end{equation}
Knowing that the particles are essentially isolated during this second
plateau, the probabilities $p_n(\infty) = \overline{p}_n$ are then:
\begin{equation}
\overline{p}_0 \simeq 0.85, \qquad \overline{p}_1 \simeq 0.15, \qquad 
\overline{p}_2 \simeq 0, \qquad \overline{p}_3 \simeq 0.
\end{equation}

\subsubsection{Third regime: activation and slow diffusion}

The third regime corresponds to the activated regime where the system
has to overcome an energy barrier to reach equilibrium, and occurs on
a timescale $t \sim e^{2 \beta}$.  An analogy in this third regime may
be drawn with a reaction--diffusion process of the type $A + A \to
\emptyset$.  When two isolated particles moving with an effective
diffusive constant $e^{-2\beta}$ through successive
creation--annihilation processes pair up, this pair diffuse on a
faster timescale ($\sim D^{-1}$) before they disappear when it reaches
another isolated particle. We therefore expect a power law decay of
the concentration during the last regime.

\begin{figure}[bt]
\centerline{\epsfig{figure=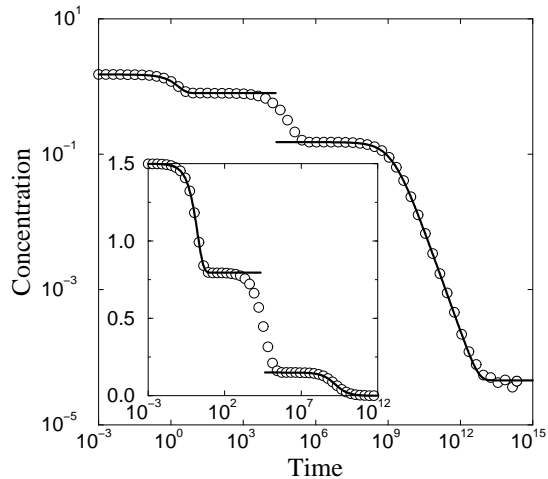,width=6.5cm,angle=-90}}
\caption{\label{decay2} Concentration decay as function of time after
a quench to the temperature $T = 1/10$ and for a diffusive constant $D
= 10^{-4}$. Symbols correspond to numerical simulations, and lines to
the analytical results for the first and third regimes. Inset:
$T=1/6$.}
\end{figure}

In order to confirm this observation we first need to determine the
probabilities $p_2$ and $p_3$ up to the order $e^{-2\beta}$,
neglecting $dp_2/dt$ and $dp_3/dt$ which are of order
$e^{-4\beta}$. From (\ref{dyn_c}) and (\ref{dyn_d}) we deduce:
\begin{equation}
\label{p3p2approx}
p_3 \simeq e^{-2 \beta} p_1, \;\;\;
D p_2 \simeq e^{-2\beta} \left(D p_0 + p_1 \right). 
\end{equation}
Replacing these expressions in (\ref{dyn_b}),
\begin{equation}
\label{p1D}
\frac{dp_1}{dt} \simeq - 2 e^{-2\beta} p_1^2 
\end{equation}
we obtain the mean--field equation for the reaction--diffusion 
process $A + A \to \emptyset$ with an effective diffusion constant 
$e^{-2\beta}$. The solution of this equation is:
\begin{equation}
\label{p1wrong}
p_1(t) \simeq \frac{\overline{p}_1}{1+2 e^{-2\beta}
\overline{p}_1 (t-\overline{t})}
\end{equation}
with $\overline{t}$ an initial time onto the second plateau 
of the concentration decay ($D^{-1} \ll \overline{t} \ll 
e^{2 \beta}$). $\overline{t} \simeq 50 D^{-1}$ is a good 
estimate for $D = 10^{-3}$ and $T = 1/10$ (see Fig.~\ref{decay2}).  

From Eq. (\ref{p1wrong}) it seems that the concentration does not
reach its equilibrium value $c_{eq} \sim e^{-\beta}$ but decays to
zero asymptotically. In order to account for the approach to
equilibrium we need to include in (\ref{p1D}) terms of order
$e^{-4\beta}$ previously neglected, which become relevant when $p_1
\sim e^{-\beta}$. Taking into account leading order terms in
$e^{-4\beta}$ (which means neglecting terms of order $p_1 e^{-4
\beta}$, which are always negligible), we obtain:
\begin{equation}
\frac{dp_1}{dt} = - 2 e^{-2\beta} \left(p_1^2 - e^{-2\beta} \right)
\end{equation}
which has for solution:
\begin{equation}
\label{p1true}
p_1(t) = e^{-\beta} \frac{\displaystyle{\overline{p}_1 + e^{-\beta} +
(\overline{p}_1 - e^{-\beta}) \, e^{ - 4 e^{-3 \beta}
(t-\overline{t})}}}{\displaystyle{\overline{p}_1 + e^{-\beta} - 
(\overline{p}_1 - e^{-\beta}) \, e^{- 4 e^{-3 \beta}
(t-\overline{t}) }}}.
\end{equation}
For a timescale $t \sim O(e^{2\beta})$ we recover the behaviour
(\ref{p1wrong}) but for a timescale $t \sim O(e^{3 \beta})$ 
the concentration reaches the equilibrium one as:
\begin{equation}
\label{p1last}
\displaystyle{c(t) \simeq p_1(t) \simeq e^{-\beta} \left( 1 + 2 e^{-4
e^{-3\beta} (t-\overline{t})} \right)  }
\end{equation}
leading to an equilibration time $\tau_{eq} \simeq e^{3 \beta}$ larger
than the relaxation time $\tau \simeq e^{2 \beta}$ deduced from
equilibrium autocorrelation.

\subsection{Out--of--equilibrium correlation and response}

We now concentrate on the behaviour of two--time quantities, like
correlation and response functions, in the out--of--equilibrium
regime.

\subsubsection{Correlation functions}

From the two--time out--of--equilibrium autocorrelation functions 
\begin{equation}
C_{n,n'}(t,t_w) = \langle \delta_{n_i(t),n} \delta_{n_i(t_w),n'} \rangle
\end{equation}
with initial conditions
\begin{equation}
C_{n,n'}(t_w,t_w) = p_n (t_w) \delta_{n,n'}
\end{equation}
it is possible to construct all relevant two point autocorrelations,
and in particular
\begin{equation}
C(t,t_w) \equiv \langle n_i(t) n_i(t_w) \rangle = \sum_{n,n'} n n' 
C_{n,n'} (t,t_w).
\end{equation}
The correlation functions $C_{n,n'}$ correspond to the probabilities
of having $n$ particles at time $t$ on a given site when there were
$n'$ particles at time $t_w \leq t$ on this particular site. They
satisfy the equations (\ref{dynamic}) replacing $p^i_n(t)$ by
$C_{n,n'}(t,t_w)$ but keeping the probabilities $p_n(t)$. This leads
to an explicitly time dependent linear system of equations for the
autocorrelations.

In the following we concentrate on waiting times $t_w \gg D^{-1}$,
that is, after the second plateau in the concentration decay. In this
case, sites have mainly at most one particle. The probabilities to
have $2$ or $3$ particles on a given site are small and satisfy the
equations (\ref{p3p2approx}). The only relevant correlation function
among $C_{n,n'}$ is $C_{1,1}$, and $C(t,t_w) \simeq C_{1,1}(t,t_w)$.
The correlation functions with $n$ or $n'$ larger than one are at
least of order $e^{-2 \beta}$ smaller, whereas those with $n=0$ or
$n'=0$ do not affect $C(t,t_w)$. A simple calculation (see appendix
\ref{app_correl}) leads the following differential equation for
$C_{1,1}$
\begin{equation}
\frac{d C_{1,1}}{dt} (t,t_w) = - C_{1,1}(t,t_w) \left[ 1 + 2 p_1 (t) +
\frac{D}{1+D} \right] e^{-2 \beta} + \left( 1 + \frac{D}{1+D} \right)
p_1 (t_w) p_1 (t) e^{-2 \beta}.
\end{equation}
with the solution
\begin{equation}
\label{correl_ooe}
C_{1,1}(t,t_w) = p_1 (t) \left[ p_1(t_w) + (1-p_1(t_w))
e^{-(t-t_w)/\tau_c} \right].
\end{equation}
The correlation time $\tau_c$ is given by:
\begin{equation}
\tau_c = e^{2\beta} \left(1 + \frac{D}{1+D} \right)^{-1}.
\label{tauc}
\end{equation}
From this solution we recover that $C_{1,1}(t,t) = p_1(t)$, which is
given by (\ref{p1true}).

\begin{figure}[bt]
\centerline{\epsfig{figure=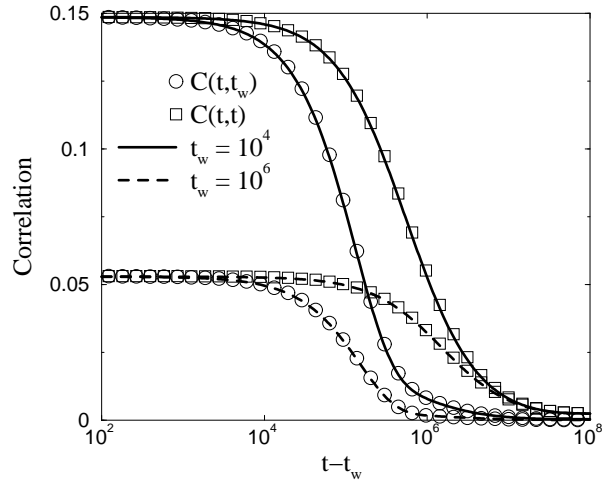,width=6.5cm,angle=-90}}
\caption{\label{correl} Out--of--equilibrium $C_{1,1}(t,t_w)$
(circles) and $C_{1,1}(t,t)$ (squares) after a quench to temperature
$T = 1/6$.  Symbols correspond to simulations and lines to the
analytical result.  Waiting times are $t_w = 10^4$ (full lines) and
$10^6$ (dashed lines), and the diffusion constant $D = 10^{-2}$.  }
\end{figure}

Fig.~\ref{correl} presents the autocorrelations $C_{1,1}(t,t)$ and
$C_{1,1}(t,t_w)$ for two different waiting times $t_w = 10^4$ and
$10^6$ as function of the time difference $t-t_w$. The temperature is
$T = 1/6$ and the diffusive constant $D = 10^{-2}$. The analytical
results Eqs.(\ref{p1true}) and (\ref{correl_ooe}) agree with the
numerical simulations for a system size $N = 10^6$.
 
\subsubsection{Response functions}

In order to determine the out--of--equilibrium response function, we
need to introduce a perturbation at time $t_w$ after the quench.
Different perturbations are possible but to get a response related to
the autocorrelation, it is common to consider a random field on each
site coupled to the corresponding observable \cite{Barrat}. The
simplest possibility is to couple the random field to the single
occupancy operator $\delta_{n_i,1}$, leading to the perturbation
\begin{equation}
\label{perturb2}
\delta H = - h \sum_i \epsilon_i \delta_{n_i,1}.
\end{equation}
$h$ is the strength of the field and will have to be small enough to
stay in the linear regime. The random variables $\epsilon_i$ may be
Gaussian variables or Ising spins ($\epsilon_i = \pm 1$) with zero
mean and unit variance. The corresponding (integrated) response
function is the change in the expectation value of $\delta_{n(t),1}$
due to the perturbation,
\begin{equation}
\label{chidef}
\chi_1(t,t_w) = \overline{h^{-1} N^{-1} \sum_i \epsilon_i
\langle \delta_{n_i(t),1} \rangle_h}
\end{equation}
where the overline stands for the average over the random field
variables. This response is conjugate to the autocorrelation
$C_{1,1}(t,t')$, which as was shown above is the relevant one for long
times and low temperatures,
\begin{equation}
\overline{N^{-1} \sum_{i,j} \epsilon_i \epsilon_j \langle \delta_{n_i(t),1} 
\delta_{n_j(t'),1} \rangle} = C_{1,1}(t,t').
\end{equation}

The definition of the perturbation is not enough to determine the
response. We also have to define how this perturbation affects the
dynamical rules, maintaining the detailed balance conditions in order
to ensure equilibrium asymptotically.  Once again different
definitions are possible. We will consider two of them. The natural
definition of the perturbed dynamics is to use for the rates a
Metropolis rule with the perturbed Hamiltonian $H+\delta H$,
\begin{equation}
\min (1,e^{-\beta \Delta (H + \delta H)}) \;\;\;\;\;\; ({\rm M})
\end{equation}
where $\Delta (H+\delta H)$ corresponds to the change in the perturbed
Hamiltonian under the corresponding transition.  One disadvantage is
that this definition will only extract a response from unoccupied
sites. A second possibility is to modify the dynamical rules by
multiplying the unperturbed rates by another Metropolis factor
\begin{equation}
\label{mm}
\min (1,e^{-\beta \Delta (\delta H)}) \times 
\min (1,e^{-\beta \Delta (H)}) \;\;\;\;\;\; ({\rm MM})
\end{equation}
It is easy to see that this modification of the dynamical rules
preserves detailed balance with respect to $H+\delta H$. The advantage
is that this definition allows to extract a response from occupied and
unoccupied sites.  For simple spin facilitated models the two dynamics
yield equivalent responses, the second one being more efficient from
the numerical point of view. We will see below that this equivalence
does not hold for the present models.

In order to determine the response function analytically we assume
that only the site $i$ is perturbed. Its probabilities $p_n^i(t)$ are
modified accordingly:
\begin{equation}
p_n^i(t) = p_n(t) + h \epsilon_i \chi_n (t,t_w)
\end{equation}
where $p_n(t)$ are the unperturbed probabilities. The equations
(\ref{dynamic}) have to be modified to take into account the change of
rates. The zeroth order in $h$ gives back the dynamical equations
(\ref{dynamic}) for the unperturbed $p_n(t)$. The first order in $h$
leads to a system of equations for the responses $\chi_n(t,t_w)$ (see
appendix \ref{app_resp}).

\begin{center}
\begin{figure}[bt]
\epsfig{figure=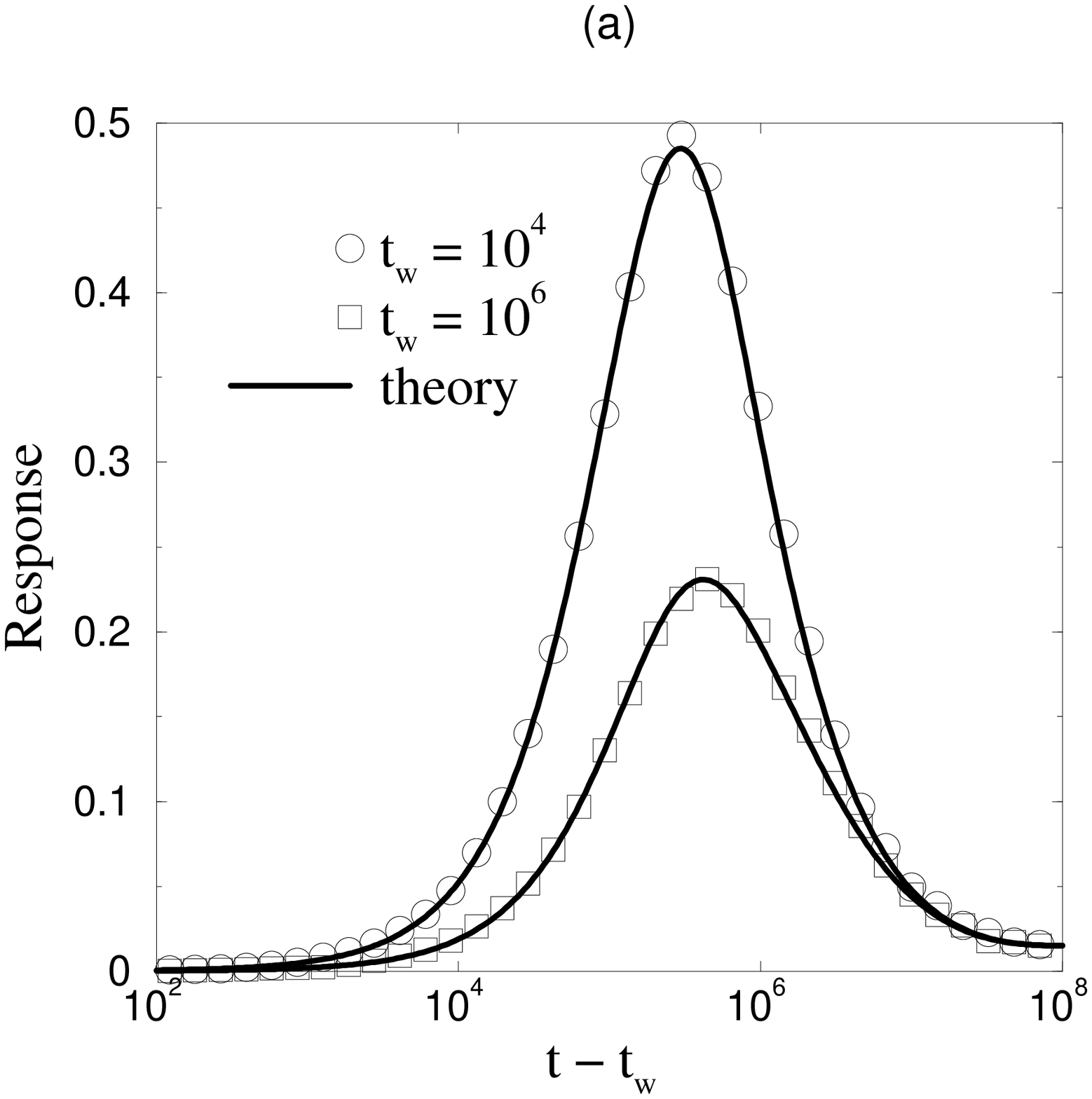,width=6.5cm,angle=-90}
\epsfig{figure=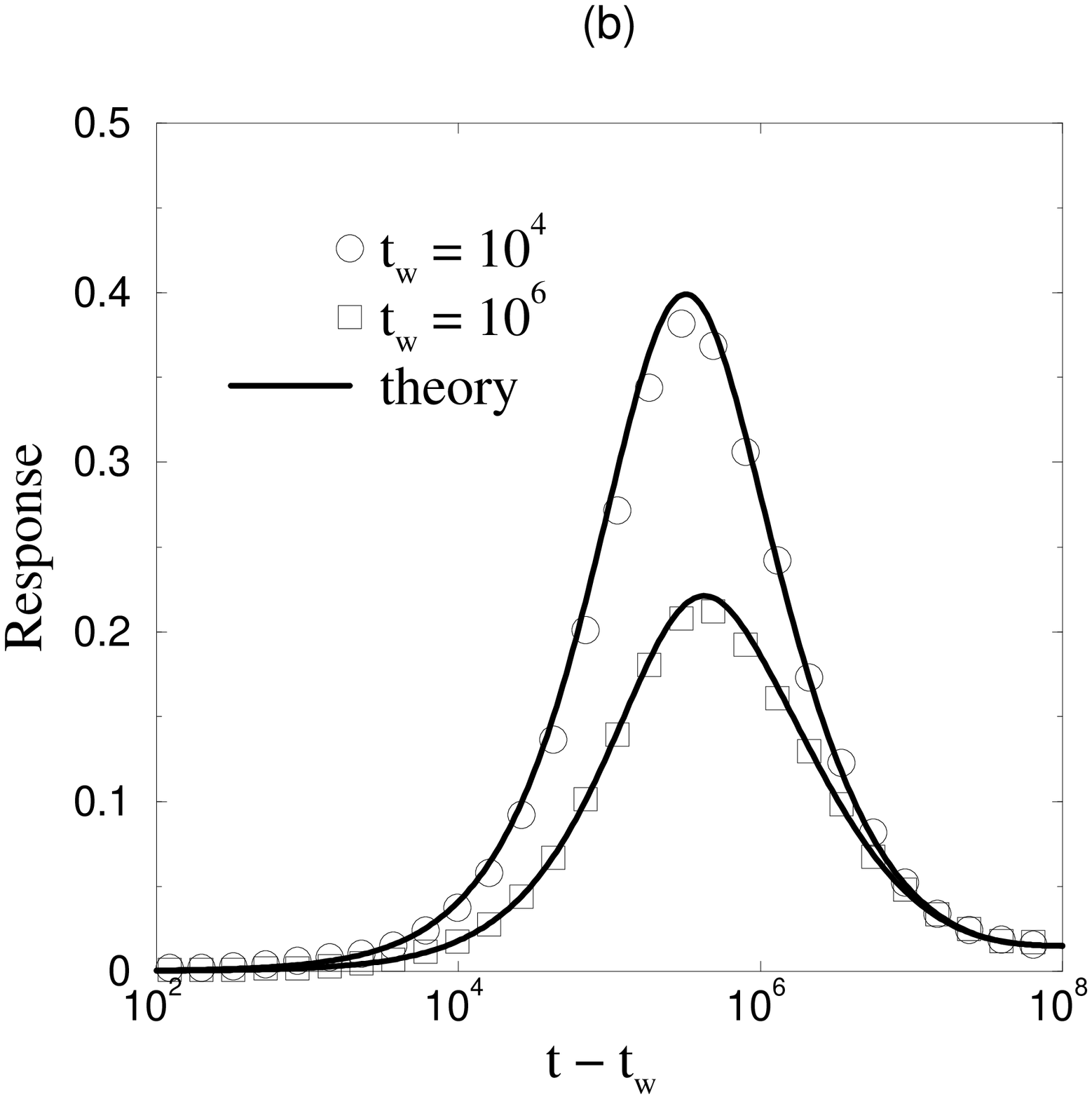,width=6.5cm,angle=-90}
\caption{\label{resp} Out--of--equilibrium response $\chi_1(t,t_w)$
for (a) MM dynamics, and (b) M dynamics, as a function of $t-t_w$, at
temperature $T = 1/6$, for waiting times $t_w = 10^4$ (circles) and
$10^6$ (squares), and a diffusive constant $D = 10^{-2}$.  The lines
correspond to the analytical result.}
\end{figure}
\end{center}

For the perturbation (\ref{perturb2}), the only relevant response
function is $\chi_1(t,t_w)$. For the case of the modified Metropolis
(MM) it satisfies (see details of the calculation in appendix
\ref{app_resp})
\begin{equation}
\frac{d}{dt} \chi_1^{\rm (MM)}(t,t_w) = - \left( 1 + 2 p_1(t) +
\frac{D}{1+D} \right) \chi_1^{\rm (MM)}(t,t_w) e^{-2\beta} + \beta
p_1(t) \left( 1 + \frac{D}{1+D} - \frac{D p_1(t)}{2(1+D)} \right)
e^{-2\beta} .
\end{equation}  
Neglecting the third term in the last parenthesis leads to the 
following approximation:
\begin{equation}
\label{respmm}
\chi_1^{\rm (MM)}(t,t_w) = \beta p_1(t) \left( 1 - e^{-(t-t_w)/\tau_c}
\right)
\end{equation}
with the time $\tau_c$ already defined for the correlation function,
Eq. (\ref{tauc}).  The timescale involved in the response and
correlation function is thus identical. Fig.~\ref{resp} shows the
accuracy of this analytic result. The response function is
non--monotonic, a common feature of the activated regime
\cite{constraints,prl}, and also observed in models of vibrated
granular matter \cite{Nicodemi,BarratLoreto}.  In the present case the
non--monotonic behaviour is given by the fact that the response is the
product of a decreasing function, $p_1(t)$, corresponding to the
number of defects able to respond, and an increasing one, $1 -
e^{-(t-t_w)/\tau_c}$, corresponding to the monotonic rescaled
equilibrium response function. The scaling from of the response
function given in (\ref{respmm}) is analogous to that in the
Fredrickson--Andersen model \cite{prl}. The corresponding calculation
for the case of the Metropolis (M) dynamics gives
\begin{equation}
\chi_1^{\rm (M)}(t,t_w) = \chi_1^{\rm (MM)}(t,t_w) - 2 \beta
\Delta(t,t_w)
\label{respm}
\end{equation}
where 
\begin{equation}
\Delta(t,t_w) \equiv p_1(t) \left[ p_1(t) - p_1(t_w) e^{-(t-t_w)/\tau_c}
\right]
\label{delta}
\end{equation}
The two responses (\ref{respmm}) and (\ref{respm}) are different.
Given that the second term inside the brackets in (\ref{delta}) decays
faster than the first one [see
Eqs. (\ref{p1wrong},\ref{p1true},\ref{p1last})], $\Delta(t,t_w)$ is
always positive, and $\chi_1^{\rm (M)} \leq \chi_1^{\rm (MM)}$, as
expected.

\subsubsection{Fluctuation--dissipation relations}

\begin{center}
\begin{figure}[bt]
\epsfig{figure=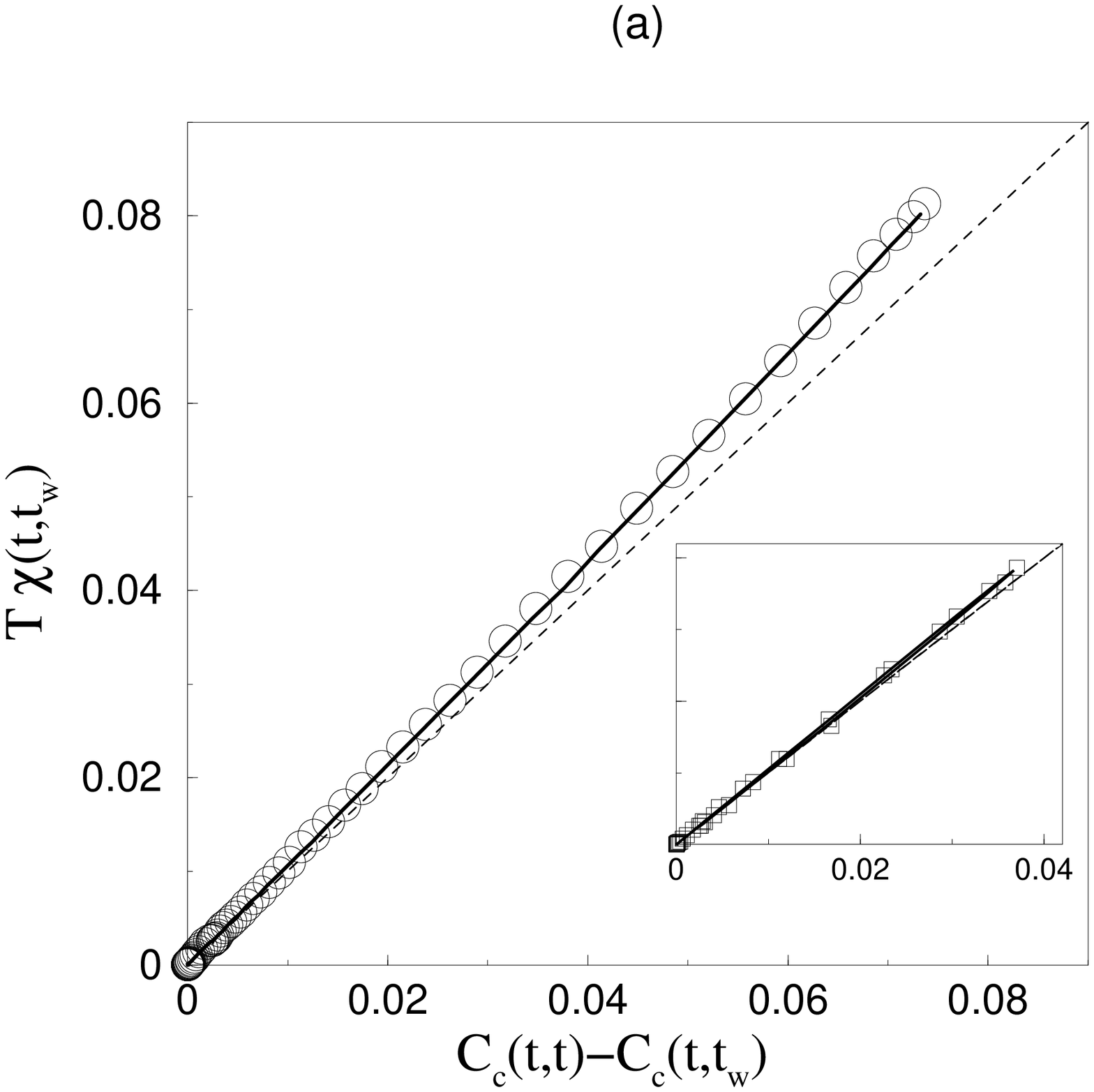,width=5.5cm,angle=-90}
\epsfig{figure=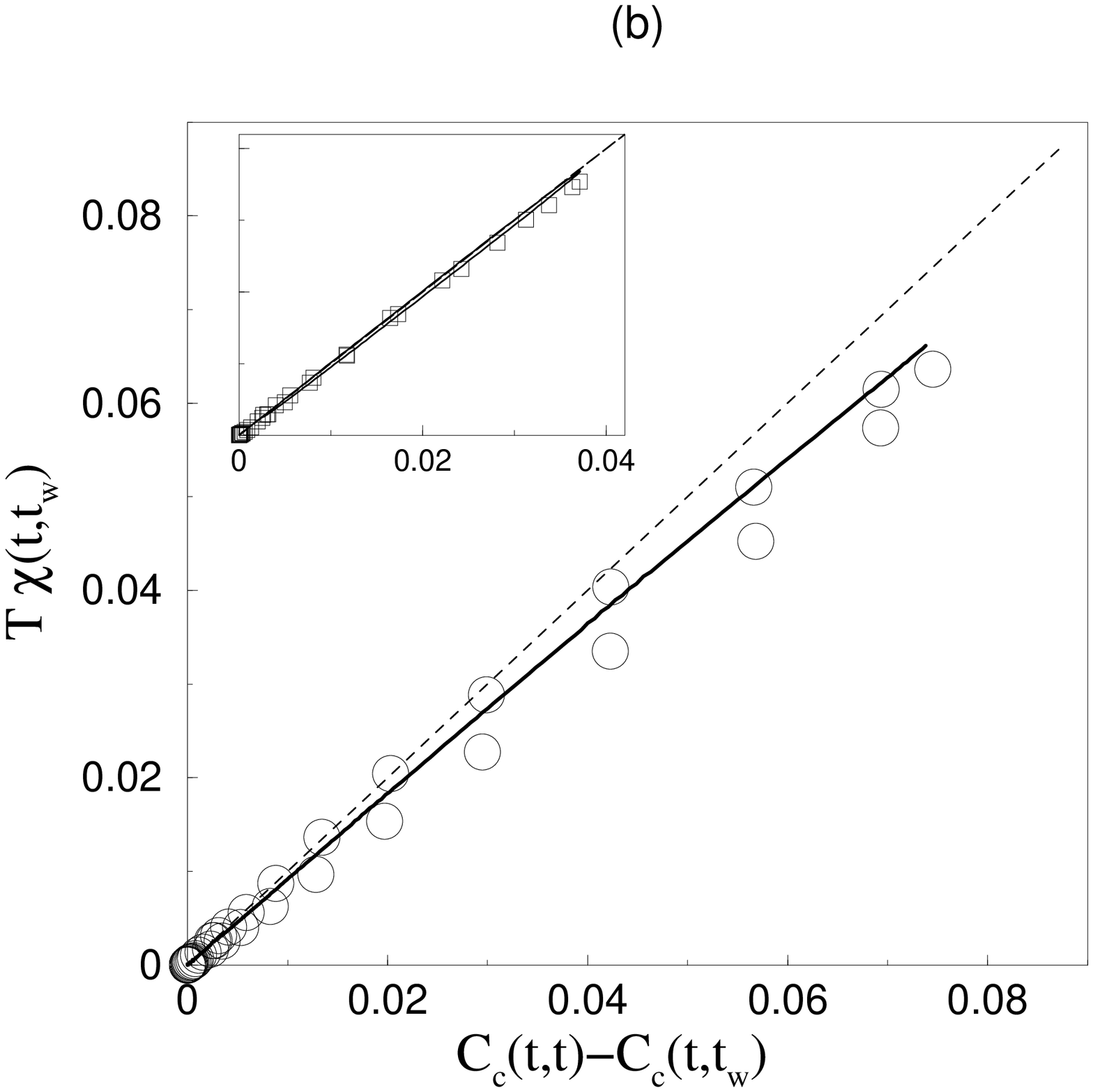,width=5.5cm,angle=-90}
\epsfig{figure=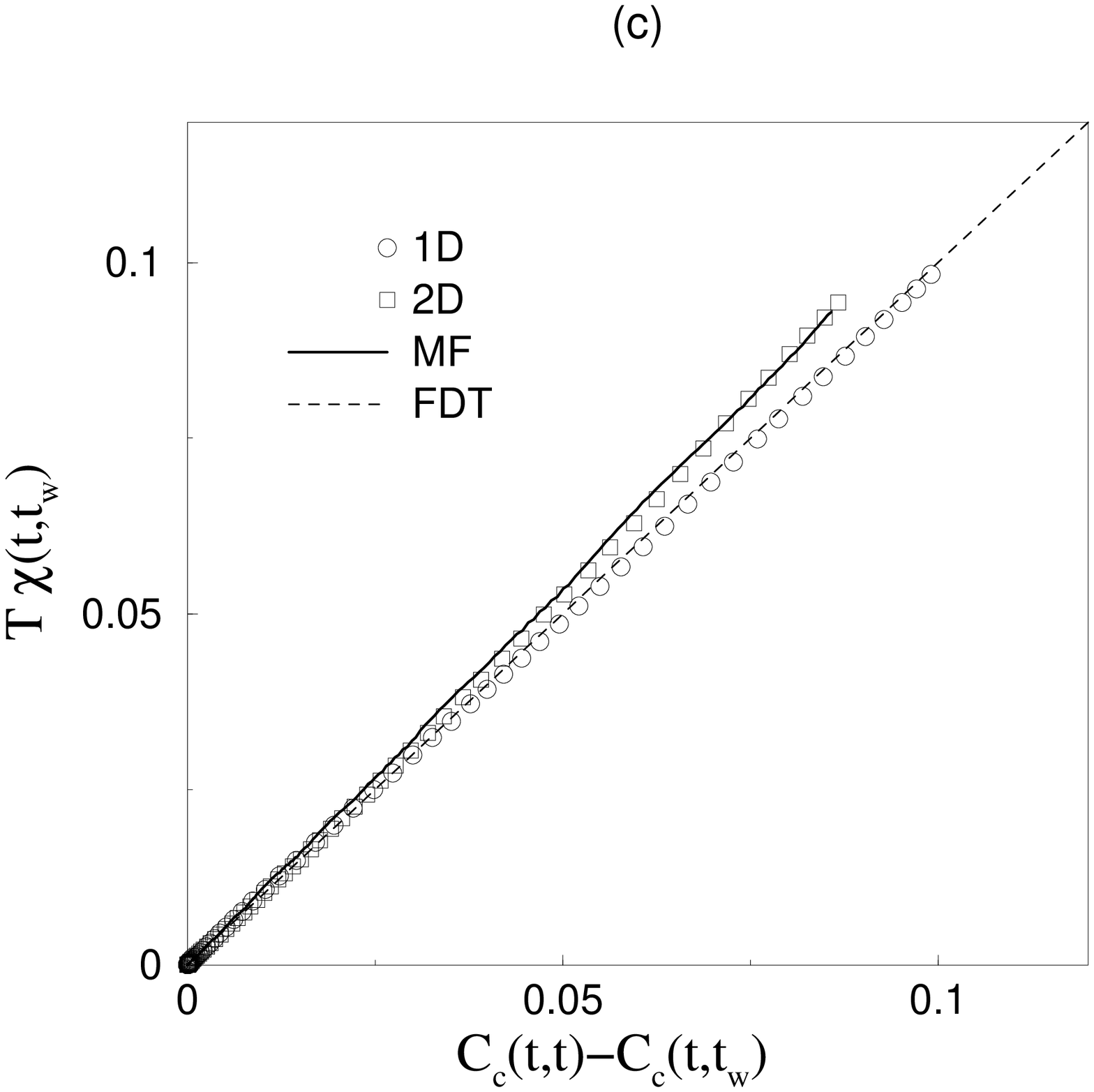,width=5.5cm,angle=-90}
\caption{\label{fdt} (a) FD plot for MM dynamics at $T=1/6$ and
waiting time $t_w = 10^4$ (inset: $t_w=10^6$).  The symbols correspond
to simulations, the full lines to the analytical result, and the
dashed line to FDT. (b) Similar plot for M dynamics. (c) FD plots in
various dimensions (for MM dynamics): $d=1$ (circles), $d=2$
(squares), and MF (full line); $T=1/6$, $t_w = 10^4$.}
\end{figure}
\end{center}

Having obtained correlation and response functions, we can now study
out--of--equilibrium fluctuation--dissipation (FD) relations
\cite{Bouchaud}. Since we are considering the case of long but finite
times, and therefore one--time quantities are still changing with
time, FD relations have to be considered between the integrated
response, $\chi_1(t,t_w)$, and the difference of the conjugate
connected correlation functions, $C_{1,1}^{(c)}(t,t) -
C_{1,1}^{(c)}(t,t_w)$, where $C_{1,1}^{(c)}(t,t') \equiv
C_{1,1}^{(c)}(t,t') - p_1(t) p_1(t')$ (see \cite{prl} for a
discussion). In Figs. \ref{fdt}(a) and \ref{fdt}(b) we show the FD
plots for the case of MM and M dynamics, respectively, for temperature
$T=1/6$ and waiting times $t_w=10^4$ and $10^6$ (inset).

There are several things to note. First, despite the fact that both
response functions and difference of connected correlations are
non--monotonic in $t$ (which implies that the FD curves when plotted
parametrically for fixed $t_w$ start from the origin, go up, and then
come back again), to a very good approximation $\chi_1(t,t_w) =
\chi_1[C_{1,1}^{(c)}(t,t) - C_{1,1}^{(c)}(t,t_w)]$, similarly to what
was found for other simple strong glass formers \cite{prl}.  Second,
the FD curves approach the fluctuation--dissipation theorem (FDT)
value as waiting time is increased, as expected. Third, the FD
relations look almost linear (although this may be just a consequence
of the fact that the departure from FDT is relatively small). In this
case the FDT violation ratio $X(t,t_w)$ \cite{cukupe,Bouchaud} is just
a function of the waiting time, $X=X(t_w)$.

We see from Figs. \ref{fdt}(a) and \ref{fdt}(b) that $X > 1$ for the
case of MM dynamics, while $X < 1$ for the case of M dynamics. This
can be traced back to
Eqs. (\ref{correl_ooe},\ref{respmm},\ref{respm}), which lead to $T
\chi_1(t,t_w) = C_{1,1}^{(c)}(t,t) - C_{1,1}^{(c)}(t,t_w) \pm
\Delta(t,t_w)$, where the upper (lower) sign corresponds to MM (M)
dynamics, together with the fact that $\Delta(t,t_w) \geq 0$ for
all times. An approximation to the value of $X(t_w)$ can be obtained
from the following argument. The slope with which the FD curves leave
the origin corresponds to the time regime in which the exponential in
the second term of Eq. (\ref{delta}) decreases much more rapidly than
$p_1(t)$ [see Eq. (\ref{p1wrong})]. If we assume that $p_1(t)$ does
not change at all in this initial period, we may approximate
$\Delta(t,t_w) \sim p_1^2(t_w) \left( 1 - e^{-(t-t_w)/\tau_c}
\right)$. This in turn gives for the FD ratio $X(t_w) \sim \left[ 1
\mp p_1(t_w) \right]^{-1}$, with the upper (lower) sign corresponding
to MM (M) dynamics. For the plots of Fig. \ref{fdt}(a) this
approximation predicts $X(t_w) \sim 1.18, 1.05$ for $t_w = 10^4,
10^6$, while a linear fit to the data gives $X_{\rm fit}(t_w) = 1.11,
1.04$ [$X(t_w) \sim 0.85, 0.95$ and $X_{\rm fit}(t_w) = 0.89, 0.96$
for Fig. \ref{fdt}(b), respectively].

Finally, in Fig.\ref{fdt}(c) we compare the behaviour in the
mean--field model with that at finite dimensions. For $d=1$ FDT is
obeyed, similar to what happens in the Fredrickson--Andersen model
\cite{prl}. For $d \geq d_c = 2$ however, the FD plots coincide with
the mean--field ones. This indicates that the aging behaviour is
controlled by the out--of--equilibrium critical point of the
underlying diffusion--annihilation process, and that mean--field
serves as a good approximation for the physically relevant dimensions
$d=2,3$.

\section{Conclusions}
\label{conclusions}

In this paper we have introduced simple models inspired by covalent
glasses and topological froths. The models display strong glass
forming behaviour despite their trivial thermodynamic properties due
to the presence of constraints to the dynamics which generate
dynamical frustration. We have studied the connection with underlying
diffusion--annihilation processes, and have shown that the aging
dynamics of these models is dominated by the critical
out--of--equilibrium fixed point of the associated
diffusion--annihilation theory.  We formulated a mean--field version
of the models, which keeps most of the features of the finite
dimensional ones, and which allows for an extremely accurate
analytical solution of the low temperature out--of--equilibrium aging
dynamics in the low temperature regime where activated processes play
a dominant role.

\acknowledgments 

This work was supported by EU Grant No.\ HPMF-CT-1999-00328, the
Glasstone Fund, and EPSRC Programme Grant GR/R83712.

\appendix

\section{First regime in the concentration decay}
\label{app_decay}

The first regime in the concentration decay may be obtained by
considering the dynamics at zero temperature and without diffusion ($D
= 0$). The dynamical equations reduce to:
\begin{eqnarray}
\frac{dp_0}{dt} & = & p_3 (p_1 + p_2)\\
\frac{dp_1}{dt} & = & p_3 (p_0 - p_1)\\
\frac{dp_2}{dt} & = & p_3 (p_1 - p_2)\\
\label{p3} 
\frac{dp_3}{dt} & = & -p_3 (p_0 + p_1)
\end{eqnarray}
with the infinite temperature probabilities $p_n = 1/4$ as initial
conditions. Notice that the right hand side of the equations is
proportional to $p_3$ and the derivative of $p_3$ is negative
indicating that a first plateau will be reached when $p_3 = 0$ with
non--zero values for the probabilities $p_0, p_1$ and $p_2$.

In the absence of diffusion, $p_2$ remains approximately constant at
its initial value, $p_2(t) \simeq 1/4$. Notice that if the initial
configuration (before the quench) does not correspond to one at
infinite temperature, this approximation is no longer valid. Using
$p_0+p_1 = 1-p_2-p_3 = 3/4 - p_3$ and (\ref{p3}), we are able to
deduce:
\begin{equation}
p_3(t) = \frac{3 e^{-3t/4}}{2 + e^{-3t/4}}.
\end{equation}
The probabilities $p_0$ and $p_1$ are determined considering the
linear combinations $a_{\pm} p_0 + p_1$ with $a_{\pm} = 
\frac12 (1\pm \sqrt{5})$ which satisfy the following differential
equation:
\begin{equation}
\frac{d}{dt} \left( a_{\pm} p_0 + p_1 \right) = \frac{p_3}{a_{\pm}}
\left( a_{\pm} p_0 + p_1 + a_{\pm}^2/4 \right).
\end{equation}
A solution is given by:
\begin{equation}
\frac{4 \left(a_{\pm} p_0(t) + p_1(t) + a_{\pm}^2/4 \right)}{a_{\pm} + 
1 + a_{\pm}^2} = \left( \frac{3}{2 + e^{-3t/4}} \right)^{1/a_{\pm}}.
\end{equation}
We deduce the following approximations:
\begin{eqnarray}
p_0(t) & = & -\frac{1}{4} + \frac{3+\sqrt{5}}{4 \sqrt{5}} 
A^{\frac{\sqrt{5}-1}{2}} - \frac{3-\sqrt{5}}{4 \sqrt{5}} 
A^{-\frac{\sqrt{5}+1}{2}}\\
p_1(t) & = & -\frac{1}{4} + \frac{1+\sqrt{5}}{4 \sqrt{5}} 
A^{\frac{\sqrt{5}-1}{2}} + \frac{\sqrt{5}-1}{4 \sqrt{5}} 
A^{-\frac{\sqrt{5}+1}{2}}
\end{eqnarray}
with $A = 3/(2+e^{-3t/4})$.

\section{Out--of--equilibrium correlation function}
\label{app_correl}

Here we determine the out--of--equilibrium correlations
$C_{n,n'}$. They satisfy the equations (\ref{dynamic}), with the
replacement of $p^i_n$ by $C_{n,n'}$.  Since all correlations with $n
\geq 2$ or $n' \geq 2$ are at least of order $e^{-2\beta}$, and the
ones with $n=0$ or $n'=0$ are irrelevant for the correlation
$C(t,t_w)$, we concentrate on the set of equations corresponding to
$n'=1$:
\begin{eqnarray}
\frac{dC_{0,1}}{dt} & = & - C_{0,1} p_3 + C_{3,1} (1-p_3) - D C_{0,1}
(p_2+p_3) + D C_{2,1} (p_0+p_1) - e^{-2\beta} C_{0,1} (1-p_0) +
e^{-2\beta} C_{1,1} p_0,\\
\label{cor_b}
\frac{dC_{1,1}}{dt} & = & - C_{1,1} p_3 + C_{0,1} p_3 - D C_{1,1}
(p_2+p_3) + D C_{3,1} (p_0+p_1) - e^{-2\beta} C_{1,1} p_0 +
e^{-2\beta} C_{2,1} p_0,\\
\label{cor_c}
\frac{dC_{2,1}}{dt} & = & - C_{2,1} p_3 + C_{1,1} p_3 - D C_{2,1}
(p_0+p_1) + D C_{0,1} (p_2+p_3) - e^{-2\beta} C_{2,1} p_0 +
e^{-2\beta} C_{3,1} p_0,\\
\label{cor_d}
\frac{dC_{3,1}}{dt} & = & - C_{3,1} (1-p_3) + C_{2,1} p_3 - D C_{3,1}
(p_0+p_1) + D C_{1,1} (p_2+p_3) - e^{-2\beta} C_{3,1} p_0 +
e^{-2\beta} C_{0,1} (1-p_0).
\end{eqnarray}
Equations (\ref{cor_c}) and (\ref{cor_d}) allow to obtain $C_{2,1}$
and $C_{3,1}$ as functions of $C_{1,1}$ and $C_{0,1}$, neglecting
their derivatives and using (\ref{p3p2approx}),
\begin{eqnarray}
D C_{2,1} & = & e^{-2\beta} \left( C_{0,1} (p_1 + D) + C_{1,1} p_1
\right), \\ (1+D) C_{3,1} & = & e^{-2\beta} \left( C_{1,1} (p_1 + D) +
C_{0,1} p_1 \right) .
\end{eqnarray}
Combining all these results and the fact that $C_{0,1}(t,t_w) +
C_{1,1}(t,t_w) \simeq p_1(t_w)$, we deduce an equation for $C_{1,1}$:
\begin{equation}
\frac{dC_{1,1}}{dt} (t,t_w) = - C_{1,1}(t,t_w) \left(1 + 2 p_1 (t) +
 \frac{D}{1+D} \right) e^{-2\beta} + \left( 1 + \frac{D}{1+D} \right)
 p_1 (t) p_1 (t_w) e^{-2\beta} .
\end{equation}
${\cal C}(t,t_w) = p_1 (t) p_1 (t_w)$ is a trivial solution of this
equation.  A general solution is then $C_{1,1}(t,t_w) = {\cal
C}(t,t_w) + \tilde{C}(t,t_w)$ with:
\begin{equation}
\frac{d\tilde{C}}{dt} (t,t_w) = - \tilde{C} (t,t_w) \left(1 + 2 p_1
(t) + \frac{D}{1+D} \right) e^{-2\beta}.
\end{equation}
The solution of this linear differential equation is:
\begin{equation}
\tilde{C} (t,t_w) = \tilde{C} \exp \left( - \frac{t-t_w}{\tau_c} -
\int_{t_w}^{t} 2 e^{-2\beta} p_1 (t') dt' \right).
\end{equation}
with $\tau_c = (1+ D/(1+D))^{-1} e^{2\beta}$. Using (\ref{p1D}), we
deduce:
\begin{equation}
\tilde{C} (t,t_w) = \hat{C} \frac{p_1(t)}{p_1(t_w)} \exp \left( -
\frac{t-t_w}{\tau_c}\right).
\end{equation}
The parameter $\hat{C}$ is obtained from the initial condition
$C_{1,1}(t_w,t_w) = p_1 (t_w)$ leading to the correlation:
\begin{equation}
C_{1,1}(t,t_w) = p_1(t) \left( p_1(t_w) + (1-p_1(t_w)) e^{ -
(t-t_w)/\tau_c} \right).
\end{equation}

\section{Out--of--equilibrium response function}
\label{app_resp}

In this appendix we determine the out--of--equilibrium response
function corresponding to a perturbation (\ref{perturb2}) in the case
of MM dynamics (\ref{mm}). This prescription leads to a different
response for spins with $\epsilon = 1$ and $-1$. The response is
determined assuming only spin $i$ is perturbed at a time $t_w$ after
the quench at $t=0$.  The corresponding probability of occupancies are
slightly modified compared to the unperturbed case:
\begin{equation}
p^i_n(t) = p_n(t) + h \epsilon_i \chi^{\epsilon_i}_n(t,t_w)
\end{equation}
which defines the two sets of response functions
$\chi^{\epsilon_i}_n(t,t_w)$ for a perturbed spin with different
random variable $\epsilon_i = \pm 1$.  The total response for a given
$n$ is simply the average between the two responses for different
$\epsilon_i$. The response functions are determined from the first
order expansion in powers of the field $h$ of the modified equations
(\ref{dynamic}). $\chi_2$ and $\chi_3$ have a higher order in
$e^{-2\beta}$ than $\chi_0$ and $\chi_1$.  From the conservation of
the probability, it follows that $\chi_0 = - \chi_1$ and we may take
$\chi_1$ as the only relevant response function.  In order to
determine $\chi_1$ we need the equations for $p^i_1$ and $p^i_3$ for a
random variable $\epsilon_i = 1$
\begin{eqnarray}
\label{dyn_b1p}
\frac{dp^i_1}{dt} & = & - e^{-\beta h} p^i_1 p_3 + p^i_0 p_3 - D
e^{-\beta h} p^i_1 (p_2+p_3) + D p^i_3 (p_0+p_1) - e^{-2\beta-\beta h}
p^i_1 p_0 + e^{-2\beta} p^i_2 p_0,\\
\label{dyn_d1p}
\frac{dp^i_3}{dt} & = & - p^i_3 (1-p_3) + p^i_2 p_3 - D p^i_3
(p_0+p_1) + D e^{-\beta h} p^i_1 (p_2+p_3) -
e^{-2\beta} p^i_3 p_0 + e^{-2\beta} p^i_0 (1-p_0),
\end{eqnarray}
while for $\epsilon_i = -1$:
\begin{eqnarray}
\label{dyn_b1m}
\frac{dp^i_1}{dt} & = & - p^i_1 p_3 + e^{-\beta h} p^i_0 p_3 - D p^i_1
(p_2+p_3) + D e^{-\beta h} p^i_3 (p_0+p_1) - e^{-2\beta} p^i_1 p_0 +
e^{-2\beta-\beta h} p^i_2 p_0,\\
\label{dyn_d1m}
\frac{dp^i_3}{dt} & = & - p^i_3 (1-p_3) + p^i_2 p_3 - D e^{-\beta h}
p^i_3 (p_0+p_1) + D p^i_1 (p_2+p_3) - e^{-2\beta} p^i_3 p_0 +
e^{-2\beta} p^i_0 (1-p_0).
\end{eqnarray}
Using (\ref{p3p2approx})  and neglecting the time
derivative of $\chi^{\epsilon}_3$, the resulting equations for
$\chi^{\epsilon_i}_n$ are:
\begin{eqnarray}
\frac{d\chi^{+}_1}{dt} & = & e^{-2\beta} \beta p_1 (1 + p_1 + D) -
e^{-2\beta} \chi^{+}_1 (1 + 2 p_1 + D) + D \chi^{+}_3,\\ \chi^{+}_3&
=& \frac{e^{-2\beta}}{1+D} \left( \chi^{+}_1 D - \beta p_1 (p_1 + D)
\right),\\ \frac{d\chi^{-}_1}{dt} & = & e^{-2\beta} \beta p_1 (p_0 +
D) - e^{-2\beta} \chi^{-}_1 (1 + 2 p_1 + D) + D \chi^{-}_3,\\
\chi^{-}_3& = & \frac{e^{-2\beta} D}{1+D} \left( \chi^{-}_1 - \beta
p_1 \right).
\end{eqnarray}
From these equations we deduce closed linear differential equations
for the response functions:
\begin{eqnarray}
\frac{d\chi^{+}_1}{dt} & = & e^{-2\beta} \beta p_1 \left(1 + \frac{p_1
+ D}{1+D} \right) - e^{-2\beta} \chi^{+}_1 \left(1 + 2 p_1 +
\frac{D}{1+D}\right),\\ \frac{d\chi^{-}_1}{dt} & = & e^{-2\beta} \beta
p_1 \left(1 - p_1 + \frac{D}{1+D} \right) - e^{-2\beta} \chi^{-}_1
\left(1 + 2 p_1 + \frac{D}{1+D}\right).
\end{eqnarray}
For systems where all sites $i$ are perturbed with uncorrelated
$\epsilon_i$, by self--averaging and linearity, the total response
$\chi(t,t_w) \simeq \chi_1 (t,t_w) = (\chi^{+}_1+\chi^{-}_1)/2$
satisfies:
\begin{eqnarray}
\frac{d\chi}{dt} (t,t_w) & = & - \left( 1 + 2 p_1(t) + \frac{D}{1+D}
\right) \chi(t,t_w) e^{-2\beta} + \beta p_1(t) \left( 1 +
\frac{D}{1+D} - \frac{D p_1(t)}{2(1+D)} \right) e^{-2\beta}.
\end{eqnarray}  
Discarding the third term in the last parenthesis allows us to solve
this differential equation. This term may be neglected due to the
small value of $p_1(t)$ for $t>t_w \gg D^{-1}$ or considering only the
case of small diffusion constant $D$. Replacing $\chi(t,t_w) = \beta
p_1(t) (\tilde{\chi} (t,t_w)+1)$ leads to the following equation for
$\tilde{\chi}$:
\begin{equation}
\frac{d\tilde{\chi}}{dt}(t,t_w) = - \frac{\tilde{\chi}(t,t_w)}{\tau_c}
\end{equation}
with $\tau_c$ the timescale already introduced for the correlation
function, Eq. (\ref{tauc}), $\tau_c = e^{2\beta} \left[1 + D/(1+D)
\right]^{-1}$.  A solution satisfying the initial condition
$\chi(t_w,t_w) = 0$ or $\tilde{\chi}(t_w,t_w) = -1$ is:
\begin{equation}
\tilde{\chi}(t,t_w)= -e^{-(t-t_w)/\tau_c}
\end{equation}
leading to the response function:
\begin{equation}
\chi^{\rm (MM)}(t,t_w)= \beta p_1 (t) \left( 1-e^{-(t-t_w)/\tau_c}
\right).
\end{equation}


\begin{thebibliography}{}
\bibitem{footnote1} We do not write explicitly the necessary sequence
around the `Feynman diagram' of the $T1$ process (see Fig.~\ref{T1}).
\bibitem{footnote2} Processes (\ref{anni}) also lead to a one-site
diffusion of the $A$ or $\bar{A}$ which remains after the
annihilation.
\bibitem{footnote3} The original model can also be described in terms
of ``spins'' $q_i = 6 - n_i$, which are conventionally called
``topological charges''.
\bibitem{footnote4} The Abelian nature of the lattice--based dynamics,
as opposed to the non--Abelian character of the foam, appears to be
irrelevant to dominant order.
\bibitem{footnote_2dimer} The annihilation of two dimers [process
(\ref{2anni})] and its reverse process, which are not essential for
the macroscopic dynamical behaviour, are not considered in this
minimalist model.
\end{thebibliography}
\end{document}